\documentclass[epj, twocolumn]{webofc}
\usepackage[varg]{}   
\usepackage{calc}
\usepackage{amsmath}
\usepackage{amssymb}
\usepackage{graphicx}
\usepackage{enumerate}

\woctitle{QENSWINS2014}
\begin{document}

\title{A design study of VOR: a versatile optimal resolution chopper spectrometer for  the ESS}
%
%

\author{P. P. Deen\inst{1,2}\fnsep\thanks{\email{pascale.deen@esss.se}} \and
        A. Vickery\inst{3,2}\fnsep\thanks{\email{anette@vickery.dk}} \and
        K, H, Andersen\inst{1,2}\fnsep\thanks{\email{ken.andersen@esss.se}} \and
        R. Hall-Wilton\inst{1,4}\fnsep\thanks{\email{richard.hall-wilton@esss.se}}
}

\institute{European Spallation Source ESS AB, Box 176, 22100 Lund, Sweden.
\and
           Niels Bohr Institute, University of Copenhagen, Denmark
             \and
           Technical University of Denmark, Department of Physics, Fysikvej, Lyngby, Denmark. 
           \and 
           Mid-Sweden University, SE-85170 Sundsvall, Sweden
          }

\abstract{%
VOR, the versatile optimal resolution chopper spectrometer, is designed to probe dynamic phenomena that are currently inaccessible for inelastic neutron scattering due to flux limitations. VOR is a short instrument by the standards of the European Spallation Source (ESS), 30.2 m moderator to sample, and provides instantaneous access to a broad dynamic range, 1 - 120 meV within each ESS period. The short instrument length combined with the long ESS pulse width enables a quadratic flux increase, even at longer wavelengths, by relaxing energy resolution from $\Delta$E/E = 1\% up to $\Delta$E/E = 7\%. This is impossible both on a long chopper spectrometer at the ESS and with instruments at short pulsed sources. In comparison to current day chopper spectrometers, VOR can offer an order of magnitude improvement in flux for equivalent energy resolutions, $\Delta$E/E = 1-3\%. Further relaxing the energy resolution enables VOR to gain an extra order of magnitude in flux. In addition, VOR has been optimised for repetition rate multiplication (RRM) and is therefore able to measure, in a single ESS period, 6 - 14 incident wavelengths, across a wavelength band of  9  \AA{} with a novel chopper configuration that transmits  all incident wavelengths with equivalent counting statistics. The characteristics of VOR make it a unique instrument with capabilities to access small, limited-lifetime samples and transient phenomena with inelastic neutron scattering. 
}

\maketitle

\section{Introduction}
\label{intro}
Direct geometry spectrometers are key workhorse neutron scattering instruments at reactor and spalllation source facilities. The scientific fields that commonly employ direct geometry spectrometers are  numerous and diverse ranging from magnetism and strongly correlated physics to soft matter and gas storage, to name a few. The strength of direct geometry spectrometers has been fully enhanced with the introduction of position sensitive detectors (PSD) on large area detectors as seen on many of the most recent chopper spectrometers from LET at the ISIS pulsed neutron source\cite{LET} to ARCS, SEQUOIA and CNCS at the Spallation Neutron Source \cite{SNS_TOFOverview},\cite{CNCS} 
(SNS) in Oak Ridge, IN5 at the Institut Laue-Langevin (ILL) \cite{OllivierIN5} and 4SEASON and AMATERAS at J-Parc, \cite{4SEASONS, Amateras}.  Four dimensional S({\bf Q}, $\omega$) maps are now routine and provide a wealth of information. However the disadvantage of neutron spectroscopy remains the weak scattering cross sections that require large sample volumes to maximise scattered flux.  The development of  the ESS opens up the possibility to improve flux profiles and create novel instrumentation. A drive towards smaller sample areas, probing transient phenomena and performing habitual in-situ measurements will  be realised on neutron inelastic spectrometers of the ESS.
An overview of VOR, a Versatile Optimal Resolution chopper spectrometer for  the ESS, is presented. The instrument parameters are optimised to maximise the flux provided by the ESS source onto a small sample area (1~$\times$1~cm$^{2}$) for a wide energy bandwidth 1~$<$~E~$<$~120~meV in a single ESS time period via repetition rate multiplication (RRM) \cite{RRM}. The instrument is therefore perfectly placed to study time-dependent phenomena on the minute to second timeframe, perform rapid identification of samples and access fundamentally new phases via high pressure or applied magnetic fields or via synthesis of new crystals typically too small for current neutron scattering studies. 

\section{Instrumental Overview}
\label{InstOverview}
\begin{figure}
\vspace{-10pt}
\centering
\includegraphics[width=0.5\textwidth]{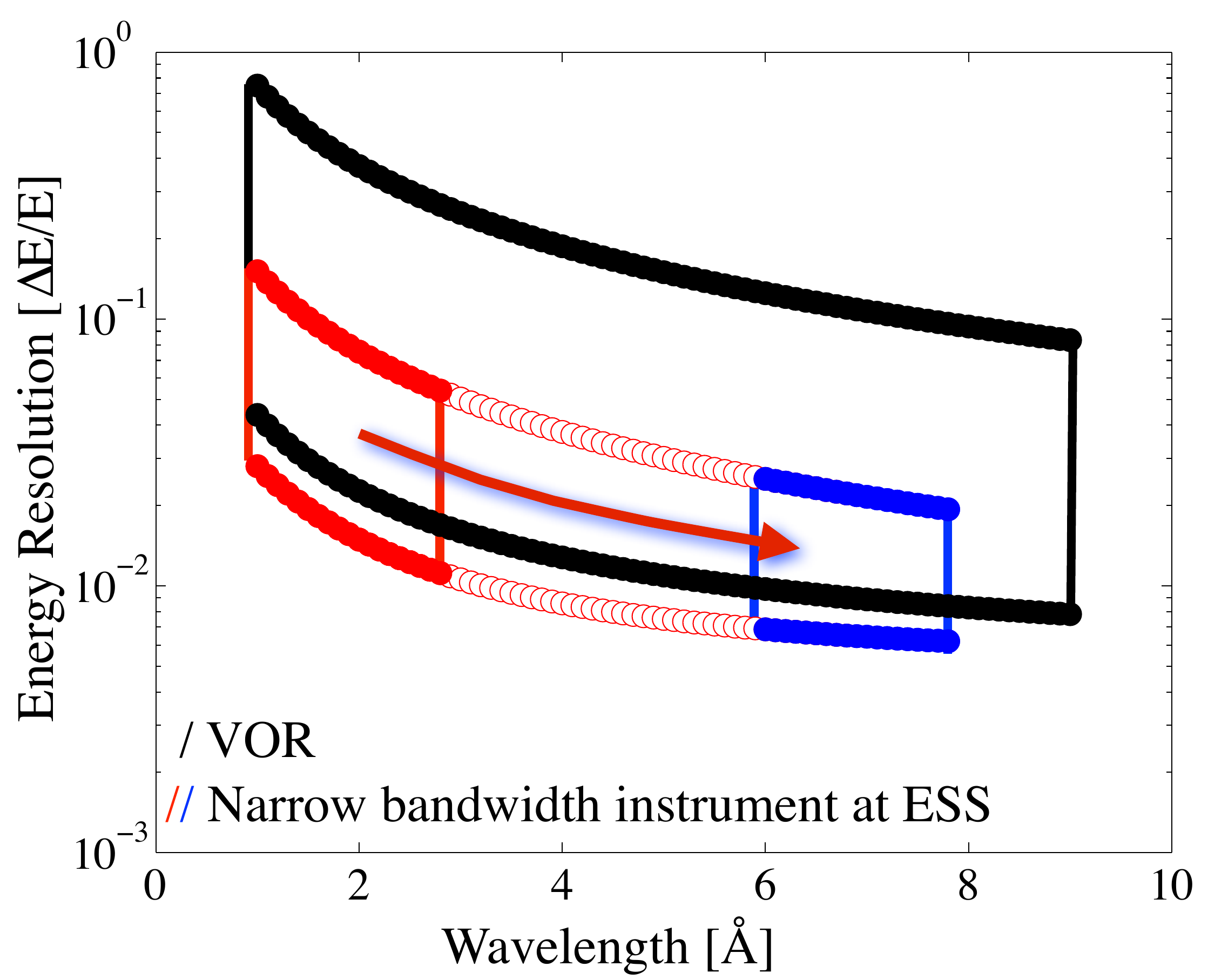}
\caption{Energy resolution-wavelength phase space probed by VOR in a single time period of the ESS in comparison to a narrow bandwidth instrument $\Delta \lambda~=~1.8$\AA{}.  The arrow indicates the region of energy resolution-wavelength that a longer narrow bandwidth instrument is able to access by shifting chopper phases.}
\label{VORPhaseSpace}
\end{figure}

VOR is a short instrument compared to most other instruments at the ESS. The moderator to sample distance is 30.2~m with a sample to detector distance of 3~m. The short instrument length directly aids the scientific profile. First, the bandwidth accessible in a single ESS pulse is given by $\textrm{tof}[\mu s] = 252.78\Delta\lambda [$\AA{}$]\textrm{L [m]}$ with L = instrument length and tof is the time of flight limiting the slowest neutrons and is the inverse of the source frequency (14 Hz),  71~ms. The bandwidth on VOR within a single ESS period is therefore  $\Delta \lambda$ $\sim$~9 \AA{}. Second, a short instrument length at a spallation source makes it possible to degrade the energy resolution and gain a quadratic increase in flux up to $\Delta E/E$ = 7~\% at the highest wavelengths. This is a consequence of the simple fact that a short instrument does not employ the full 2.86~ms ESS pulse when the high resolution is required yet can do so when flux is required at the expense of energy resolution. VOR's coverage of incident wavelength and energy resolution $\Delta E/E$, probed in a single ESS time period, via RRM \cite{RRM},  is shown in Figure \ref{VORPhaseSpace}. A comparison with a 160 m narrow bandwidth instrument , $\Delta \lambda ~=~1.8~$\AA{}, is provided. VOR is able to probe weak scattering features while accessing huge swathes of S(Q, $\omega$) to address the scientific needs outlined in the introduction. It should be noted however that the extension of phase space up to  $\Delta$E/E = 10~\% on VOR  is purely a consequence of the 2.86~ms ESS source pulse. Although  $\Delta$E/E = 10~\% is of no great use for many experiments, extending up to 5-6~\% at the expense of flux will be essential for science that is currently inaccessible. The following outline of the instrument concept for VOR is based on analytical considerations and instrument simulations using the ray tracing simulation package McStas version 2.0 \cite{willendrup2014a}.

\subsection{Beam Extraction}
\label{BeamExtraction}
\begin{figure}[htp!!]
\vspace{-10pt}
\centering
     \includegraphics[width = 0.5\textwidth]{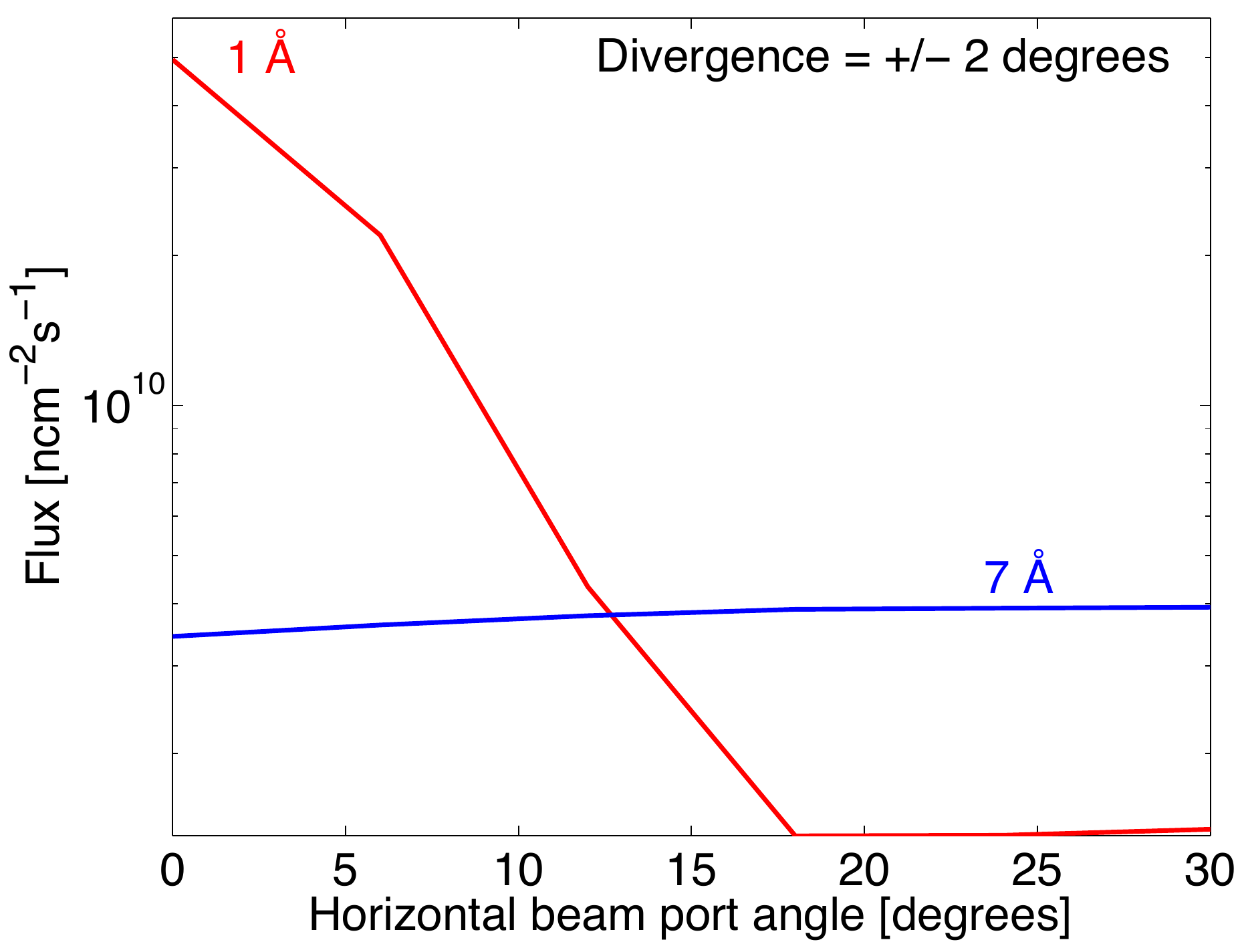}
\caption{Flux variation close to the entrance of the guide for thermal (1 \AA{}) and cold (7 \AA{}) neutrons as a function of horizontal beam port angle}
\label{VORBrightness}
\end{figure}

Beam extraction from the moderator has been optimised to extend the spectral range  of the instrument from the cold into the thermal regime. The upper energy limit is not defined by the scientific case but by the technical limits of the choppers and guide transmission.  The spatial variation of the moderator spectrum is taken advantage of to extract simultaneously cold and thermal neutrons \cite{Troels}.  Neutrons emitted from the cold and thermal moderators and from the sides of the thermal reflectors arrive at a  60$^{\circ}$ beam extraction window. Such a  60$^{\circ}$ beam window is subdivided into twelve 5$^{\circ}$ beam ports, most of which will host an instrument.  At the central port the cold flux is optimised with minimal thermal flux, see Figure \ref{VORBrightness} for the angular dependence of the 7 \AA{} cold neutron flux . However at the edge of  the 60$^{\circ}$ beam extraction window the thermal flux can be fully accessed with only a 15\% loss in cold flux,  see Figure \ref{VORBrightness} for  the 1 \AA{} thermal neutron flux . A divergence window of  $\pm$ 2$^{\circ}$ is considered. The angular dependence of other wavelengths can be extrapolated from the 1 and 7 \AA{} dependence. 

\subsection{Guide}
\begin{figure}
\centering
\includegraphics[width=0.5\textwidth]{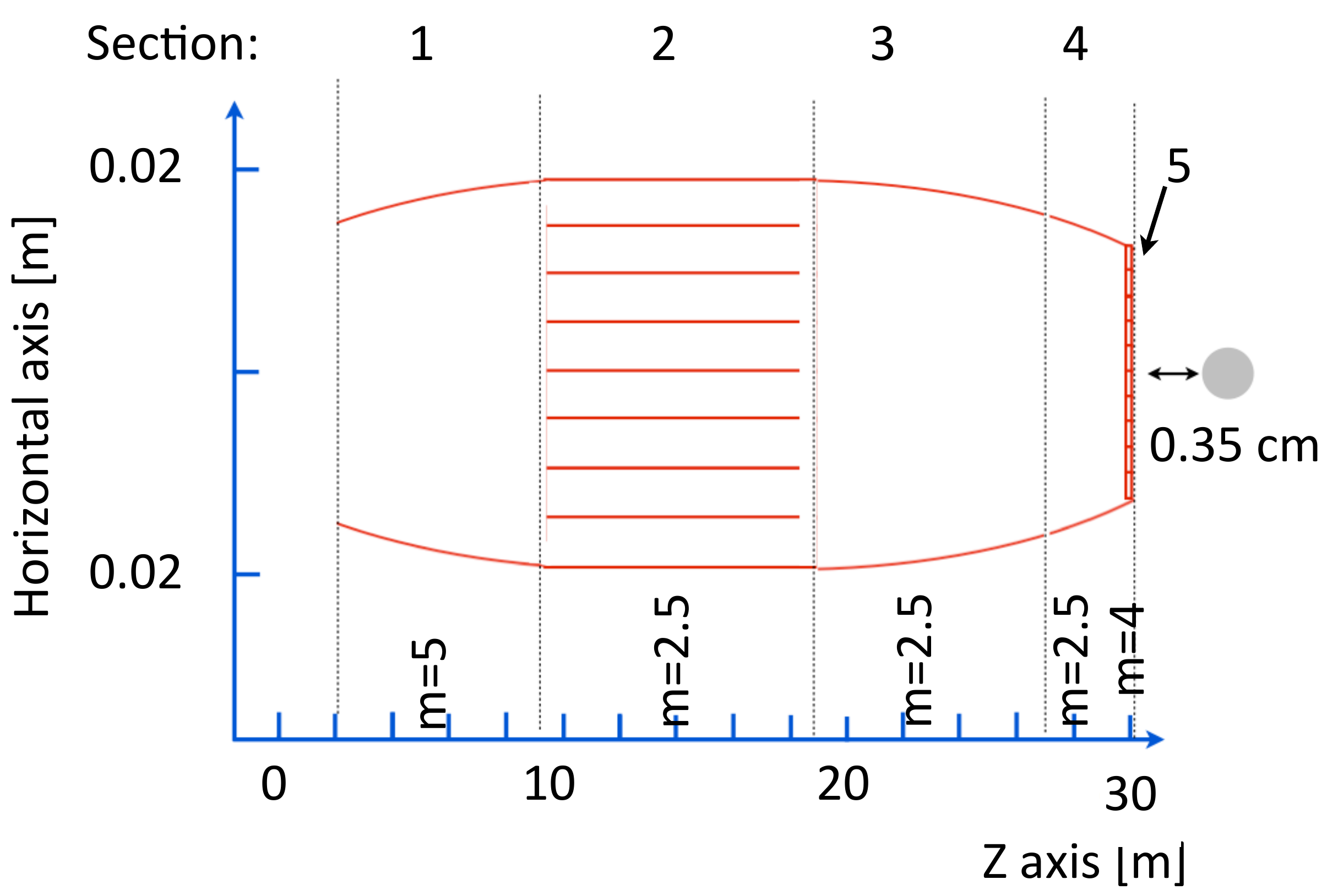}
\caption{Guide overview along the horizontal plane. The bender portion, section 2, is shown as a straight section for simplicity. The various sections are outlined in the text.}
\label{Guide}       
\end{figure}

VOR's guide aims to maximise flux for 1~$<$~$\lambda$~$<$~$9$ \AA{} onto a 1~$\times$~1~cm$^{2}$ sample area within a single ESS period of 71~ms. The acceptance of the neutron guide must ensure the transmission of a maximum divergence that will provide a $\Delta$ Q = 0.02~\AA$^{-1}$ for single crystal measurements and yet a much broader Q resolution for diffuse scattering phenomena. At 1~\AA{} the guide transports at least $\pm$0.5$^{\circ}$ while for the highest $\lambda$ the divergence profile extends across at least $\pm$2$^{\circ}$  in both the horizontal and vertical directions. For the highest {\bf Q}-resolutions the divergence profiles should result in smooth Bragg peaks at the detector to enable precise determination of lineshapes and broadening effects. To enable quantitative analysis it is important that the flux profile is uniform within~10~\% across the 1~$\times$~1~cm$^{2}$ sample profile.  \\
The guide has elliptical profiles in both the horizontal and vertical directions and can be subdivided into 5 sections, see Figure \ref{Guide}.  The focussing parameters at the start and end of each section are provided in Table \ref{GuideTable}. Section 1 extends from 2~m beyond the moderator to the first choppers just before 9.5~m. Section 2 is an 8 channel bender with parallel curved sidewalls and a 900~m radius designed to position the sample out of line of sight of the moderator. The curvature of the guide is not shown in Figure \ref{Guide} for ease of viewing. Section 3 extends from a second set of choppers at 19~m to the monochromating choppers at 28.5~m. The penultimate section, section 4, extends from the monochromating choppers to an exchangeable collimation piece that begins at 29.85~m. An exchangeable guide piece, 0.5~m,  to introduce beam polarisation can be placed in section 4. Section 5 is a 0.1~m Soller collimator. The angular acceptance of the collimator will be varied by exchanging collimator pieces. The sample position sits at 0.35~m from the guide exit. All guide simulations shown incorporate at least a 0.1~m guide gap at the chopper positions. The focussing parameters in Table \ref{GuideTable} have been optimised to ensure a small beamspot, 1~$\times$~1 cm$^{2}$, at the sample position. 
\begin{table}
\begin{minipage}[b]{0.45\textwidth}
\begin{tabular}{|c|p{0.6cm}|p{1.2cm}|p{1.2cm}|p{1.2cm}|p{1.2cm}|p{1.2cm}|}
\hline
  Sec: & L   & V-Foc. in &  V-Foc. out& H-Foc. in&  H-Foc. out\\ 
  &                      [m] &  [m] & [m] & [m]  &  [m]  \\
  \hline
  1  & 7.5 & 2 & 0.7 & 2 & 4 \\
2  & 9.5 & N/A & N/A & N/A & N/A  \\
  3  & 9.5 & 2 & 0.7 & 2 & 1.2  \\
   4  & 1.35 & 2 & 0.7 & 2 & 1.2  \\
    5  & 0.1 & N/A & N/A & N/A & N/A  \\
\hline
\end{tabular}
\caption{Parameters pertaining to the guide concept. The focussing parameters are distances from the end of the guides. For sections 2 and 5 the parameter values are not applicable (N/A) since these portions are not elliptical. }
\label{GuideTable}
\end{minipage}
\end{table}\\

\begin{figure}[h!!]
  \centering
\includegraphics[width=0.2\textwidth]{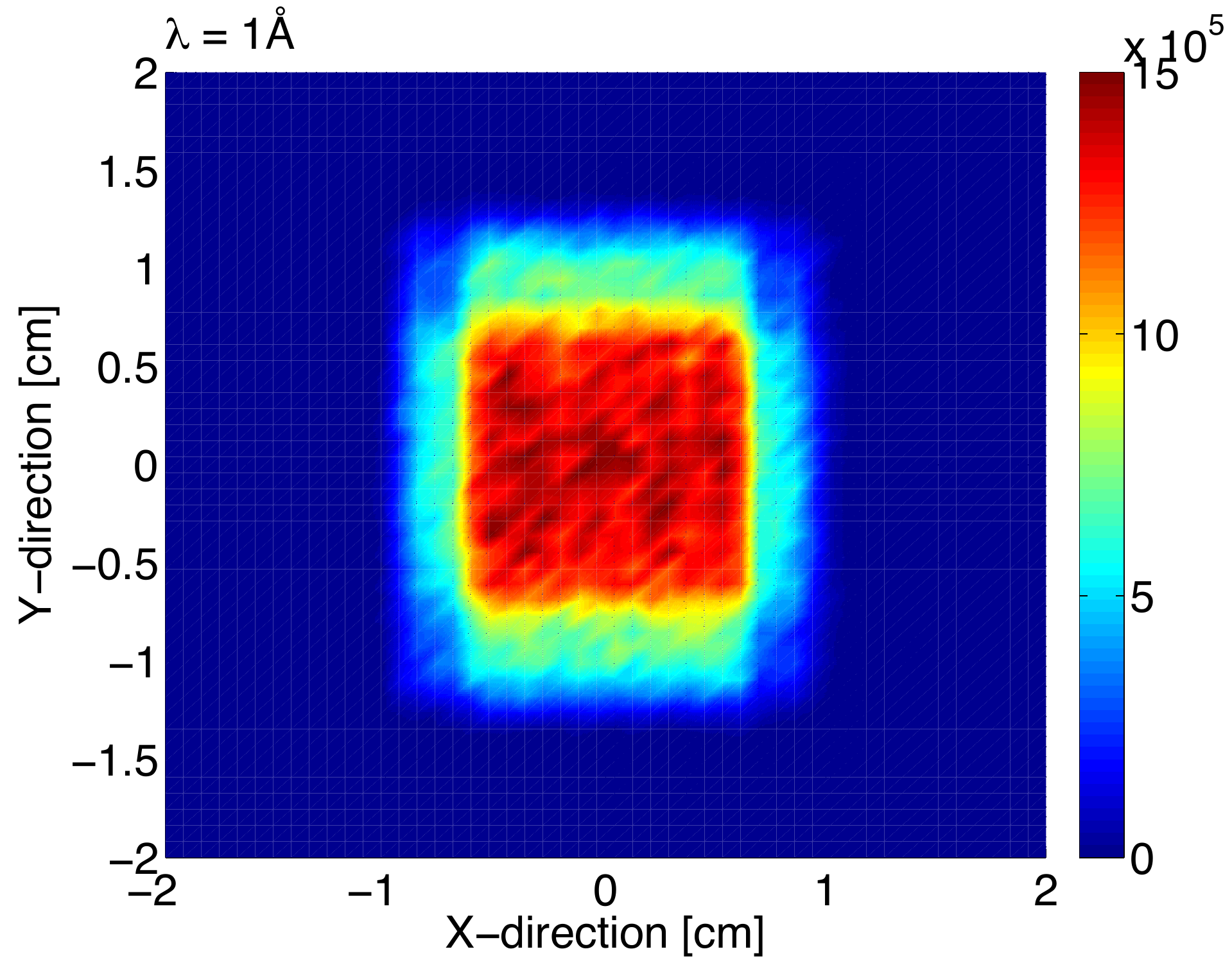}
\includegraphics[width=0.2\textwidth]{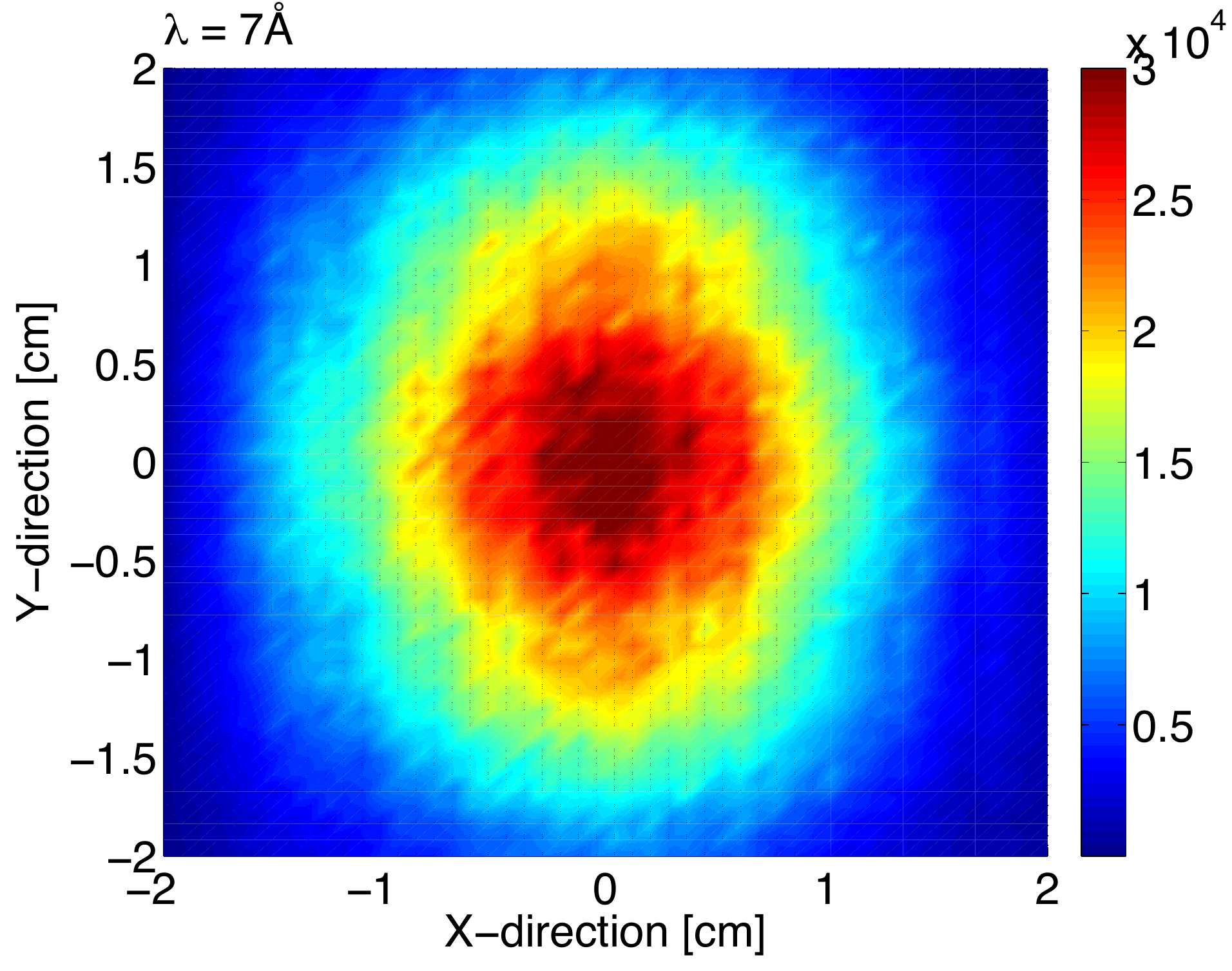}
\includegraphics[width=0.2\textwidth]{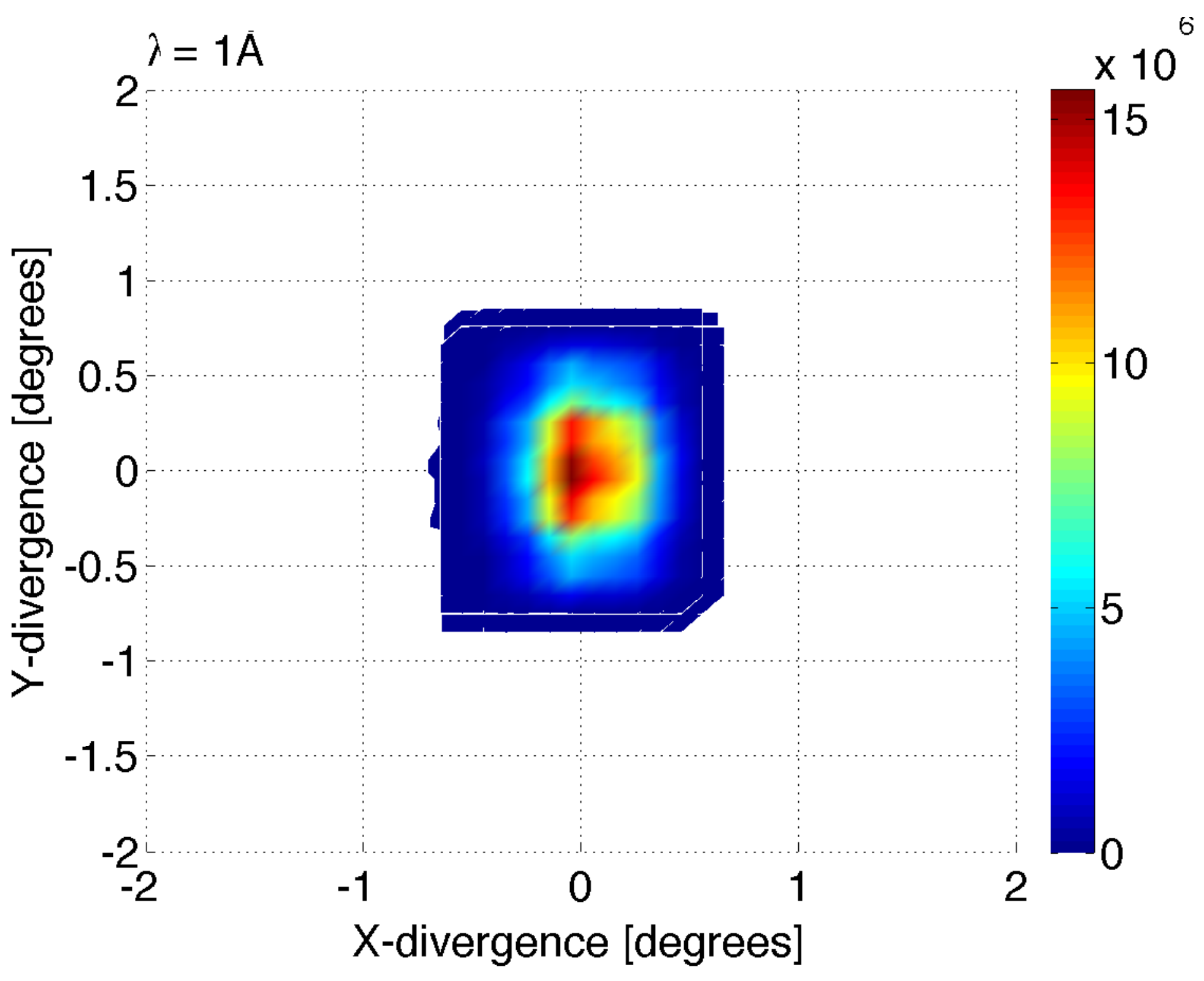}
\includegraphics[width=0.2\textwidth]{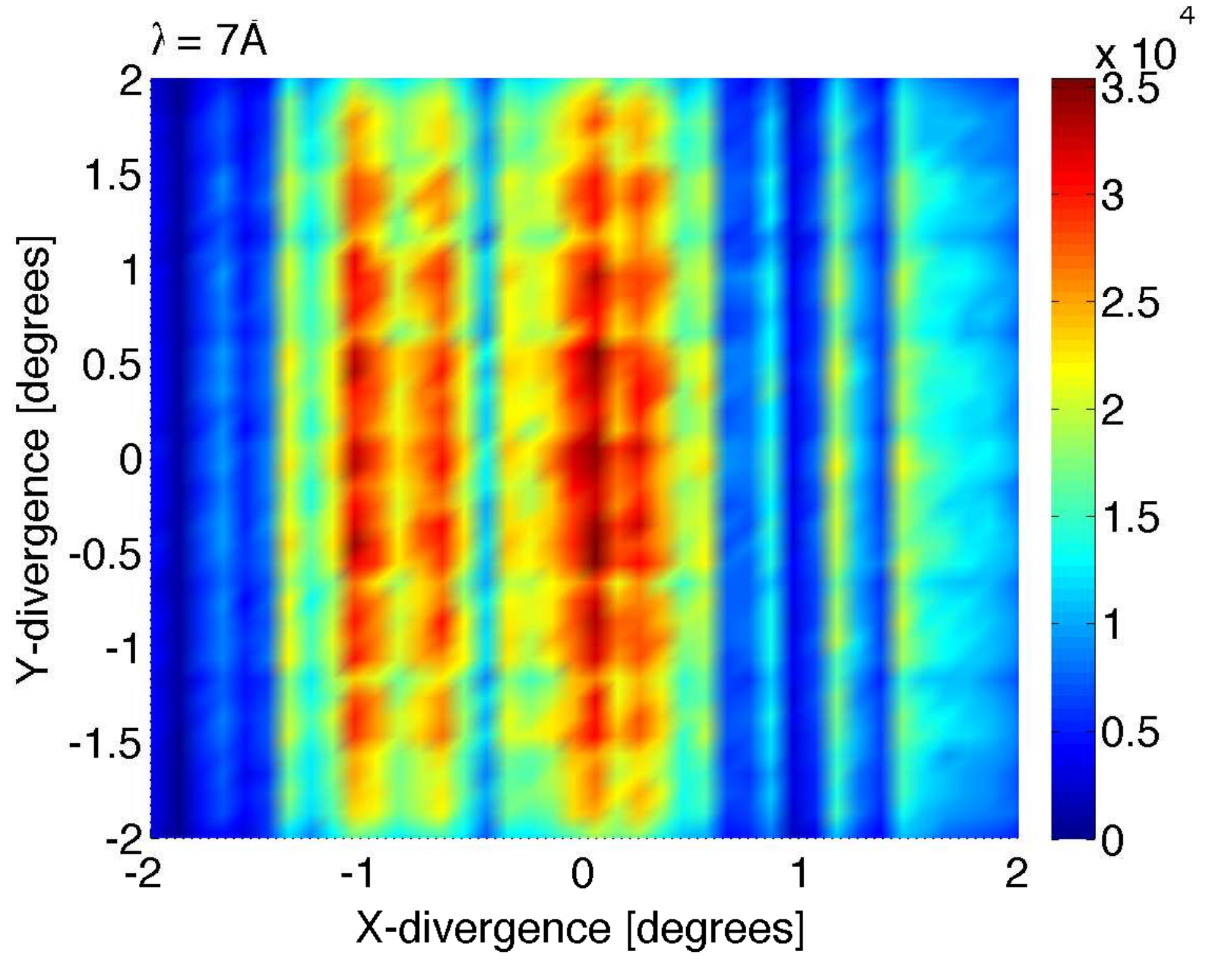}
\caption{{\small (top) Flux at the sample position for  (left) 1 \AA{} and (right) 7 \AA{}. (bottom) Divergence at the sample position for (left) 1 \AA{} and (right) 7 \AA{}.}}
\label{FluxDiv2D}
\end{figure}

\begin{figure}[htp]
\centering
\includegraphics[width=0.45\textwidth]{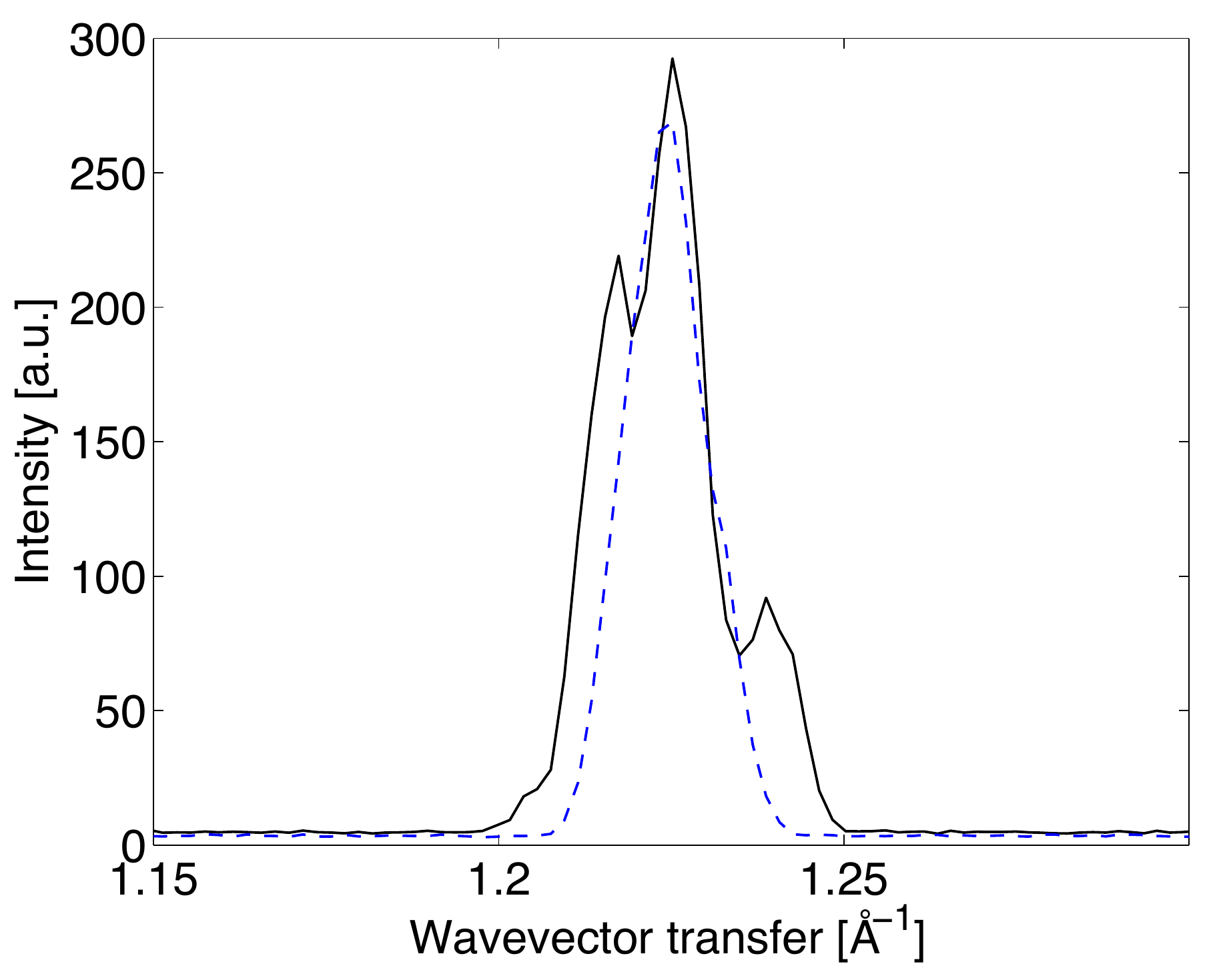}
\caption{Typical Q-dependence of powdered Na$_{2}$Ca$_{3}$Al$_{2}$F$_{14}$ with a single collimation channel (-) and 0.7$^{\circ}$ collimation (-  -) with incident $\lambda$  = 7 \AA{}. A single Bragg peak is shown for clarity.}
\label{CollimationQDep}
\end{figure}

\vspace{0.5cm}
Figure \ref{FluxDiv2D} shows the flux and divergence profile for $\lambda$~=~1 \AA{} and 7~\AA{}. The flux profile is uniform and is focussed to 1~$\times~$1 cm$^{2}$ across the complete bandwidth. The divergence profile 
 is less uniform at long wavelengths and results in Bragg diffraction as shown in Figure \ref{CollimationQDep}. Bragg scattering from a powdered Na$_{2}$Ca$_{3}$Al$_{2}$F$_{14}$ sample clearly shows the discrete features of the divergence profile without collimation, see Figure \ref{CollimationQDep}. When measuring broad, short-range order this profile would not affect the resultant scattering. However for long-range correlated features it is important to provide a clean and uniform profile. Collimating the incident beam with 0.7$^{\circ}$ collimation provides such a profile with minimal flux loss. 



\subsection{Chopper Cascade}
\begin{figure}
\centering
\includegraphics[width=0.32\textwidth]{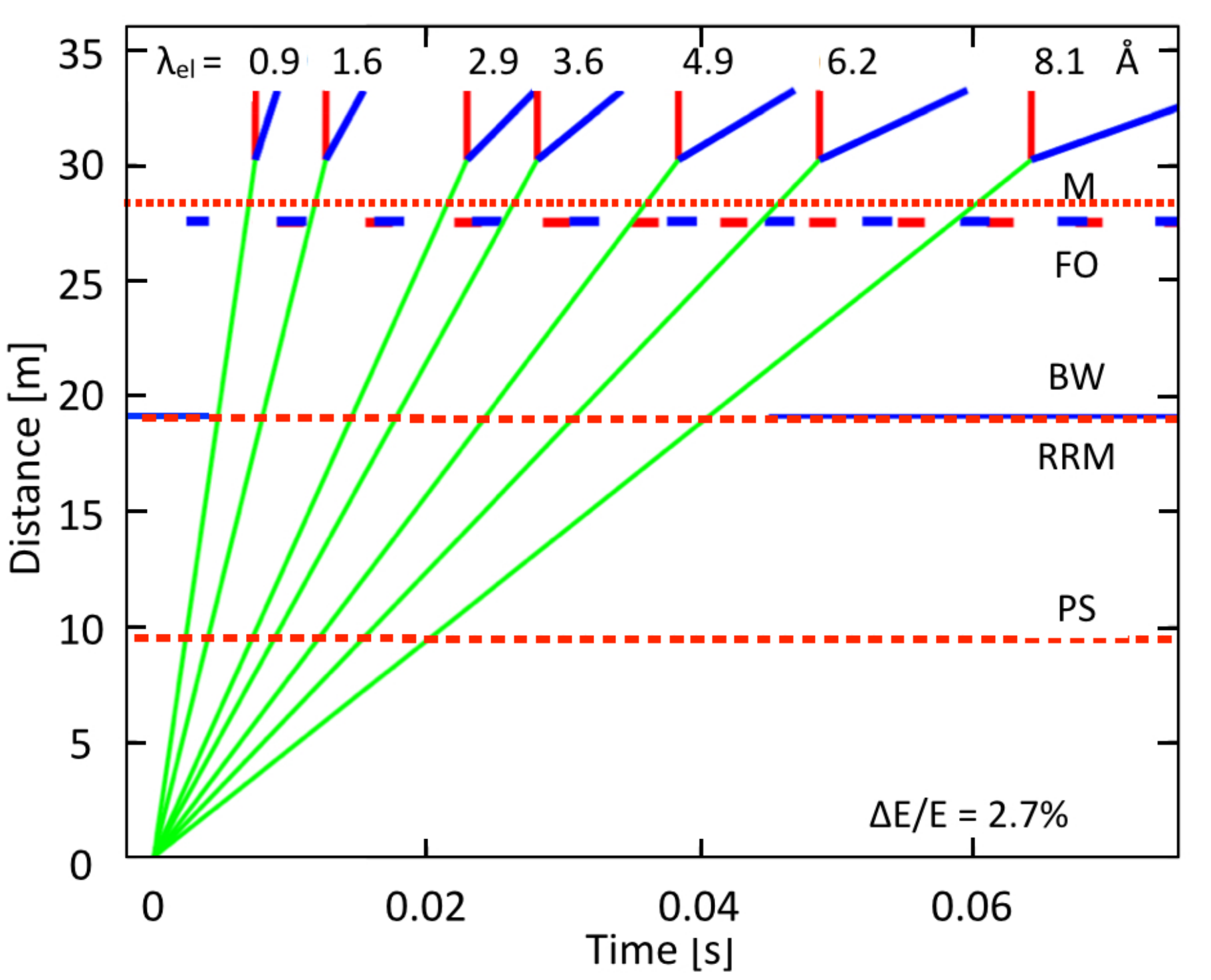}
     \includegraphics[width=0.3\textwidth]{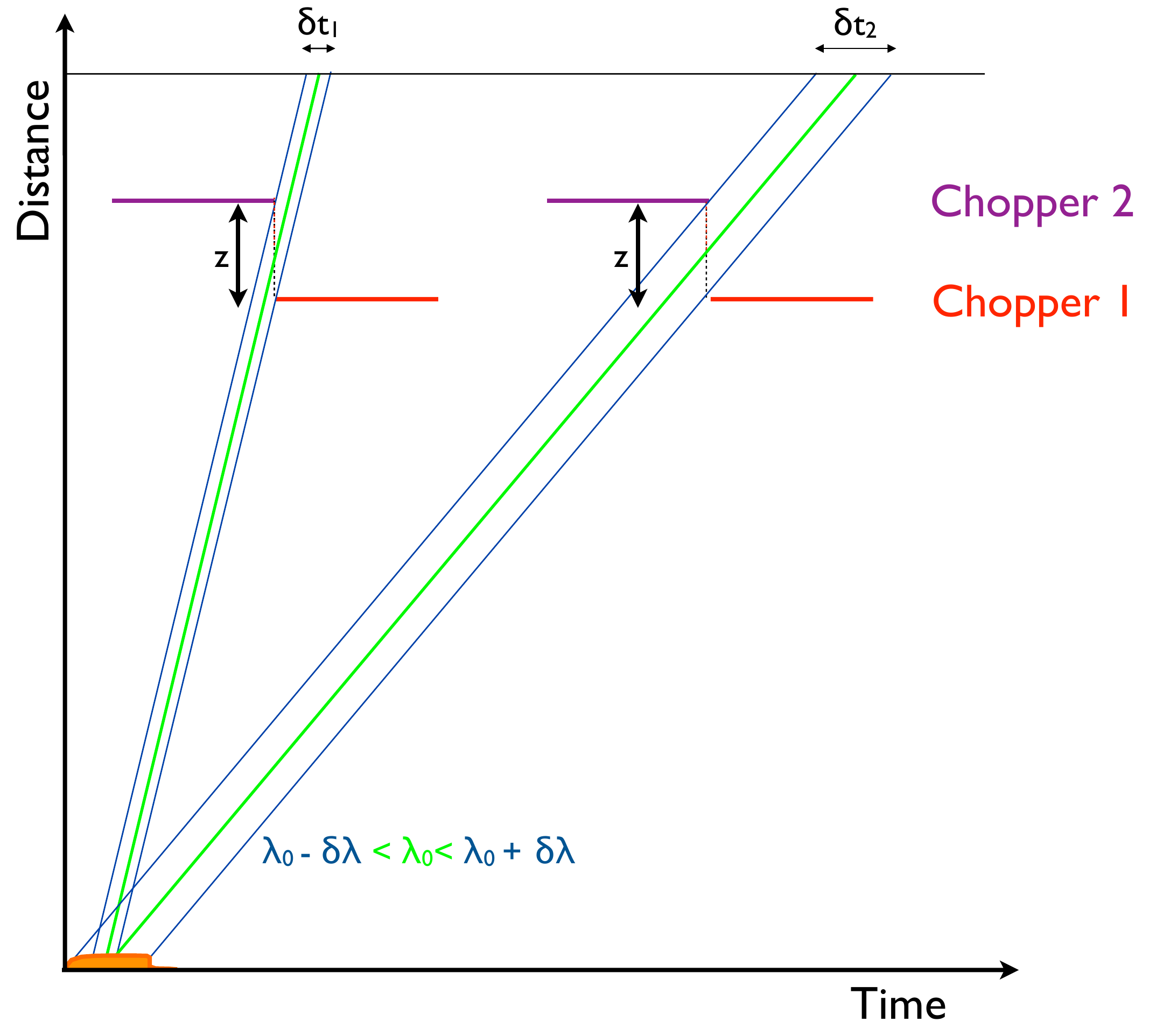}
 \includegraphics[width=0.35\textwidth]{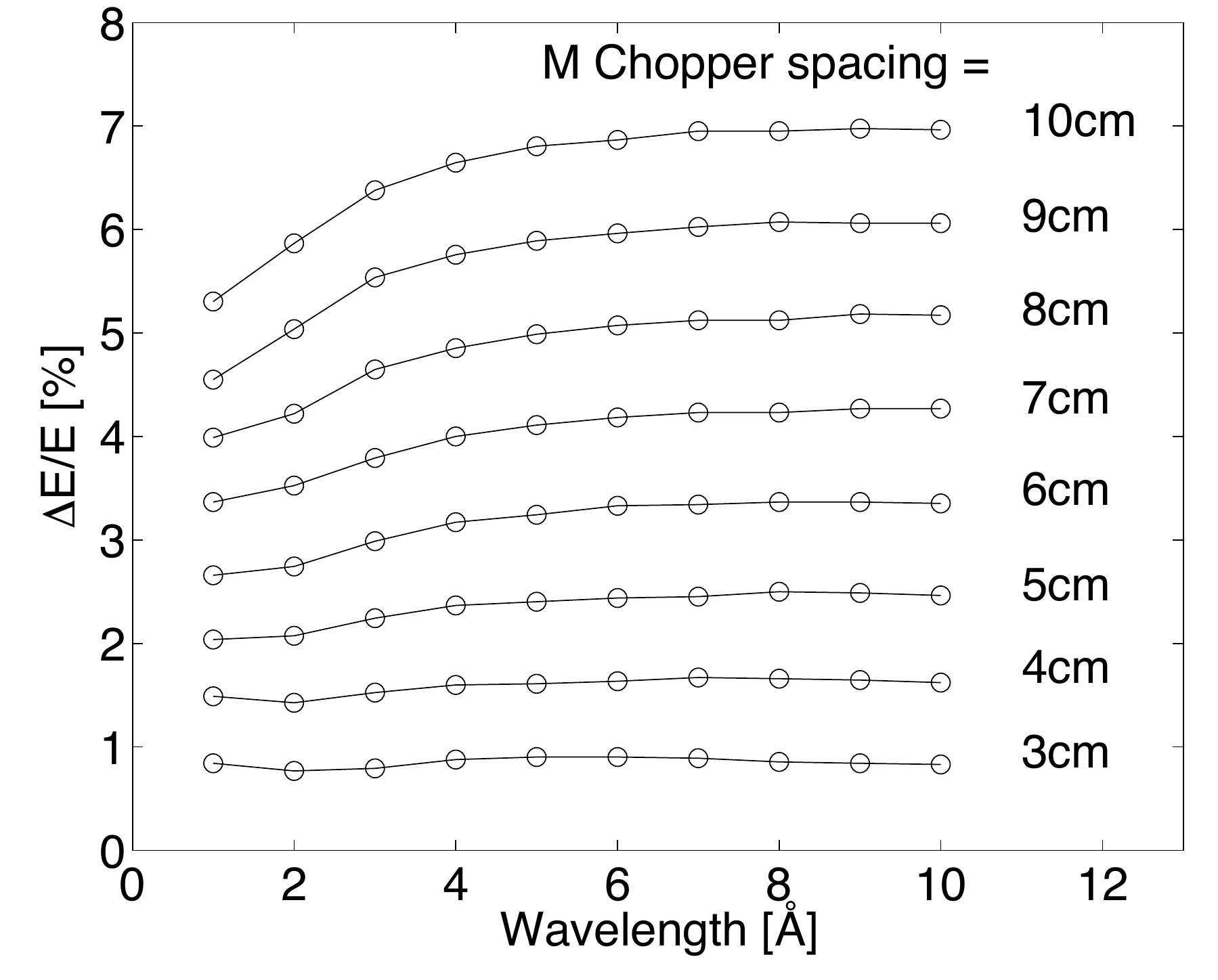}
\caption{(top)An analytical representation of the chopper cascade for an energy resolution of $\Delta$E/E = 2.7 \%. The time distance diagrams shows  the positions of the various choppers on VOR.   (middle)  Diagram outlining the double-blind configuration for the pulse shaping and monochromating choppers to achieve equivalent energy resolutions for a range wavelengths.  (bottom) McStas simulated energy resolution, $\Delta E/E$ as a function of the gap between the double-blind chopper blades at the M chopper position. An almost uniform energy resolution across a broad wavelength band is obtained.}
\label{ChopCascade}       
\end{figure}

The chopper cascade envisaged for VOR is presented in the time-distance diagram of Figure \ref{ChopCascade} with the main parameters given in Table \ref{Chopper}. The main choppers are the pulse shaping(PS), the repetition rate multiplication(RRM) and the monochromating(M) choppers positioned at x:2x:3x with x = 9.5 m.  A bandwidth (BW) chopper is necessary to separate the ESS time periods. The chopper configuration is based on the concept of balanced resolution first defined by Lechner \cite{Lechner}. To fully utilise RRM on a wide bandwidth instrument it is important to provide two aspects that are lacking on current day chopper spectrometers: (1) be able to avoid frame overlap in particular for the coldest neutron pulses and (2) provide a burst time that is optimised for each incident wavelength and resolution across the complete wavelength band. First, pulses that result in frame overlap are suppressed  using two frame overlap (FO) choppers rotating at N$\times$14~Hz and (N+1)$\times$1~4Hz, 14 Hz being the ESS source frequency and N = integer number. Rotating two chopper blades at slightly different frequencies allows their closed time windows to overlap progressively as time elapses so that they will suppress several pulses at the highest wavelengths. This is exemplified in the analytical calculations represented in Figure \ref{ChopCascade}(top). Second, in order to provide optimised burst times a chopper configuration based on the  double-blind chopper system \cite{DoubleBlindA} will be employed, see Figure \ref{ChopCascade}(middle). In a  double-blind chopper system the opening and closing edge of the time windows of two chopper disks coincide in time. Pulses with different wavelengths arriving from the ESS pulse between 0 $<$ time $<$ 2.86~ms will therefore cross the chopper windows with a burst time which is proportional to the wavelength, see Figure \ref{ChopCascade}(middle). To vary the energy resolution the burst times must be altered and this is possible by increasing or decreasing the distance between the chopper disks, distance $z$ on Figure \ref{ChopCascade}(middle). McStas simulations show that almost equivalent relative energy resolutions,  $\Delta E/E$, are achieved across VOR's entire bandwidth for a broad range of $\Delta E/E$.  Equivalent relative energy resolution across the whole bandwidth allows the user to probe the whole dynamic range with equivalent counting statistics.  In conjunction with improved fluxes provided by the ESS source brightness it will be possible to probe time-dependent phenomena down to the second timescale across a very broad dynamic range. This chopper configuration has been named the relative constant energy resolution chopper configuration (RC). The full details of the RC chopper configuration will be published elsewhere \cite{VORChoppers}. 
\vspace{1cm}

\begin{table}
\begin{minipage}[b]{0.45\textwidth}
\begin{tabular}{|p{0.7cm}|p{0.6cm}|p{1.2cm}|p{0.6cm}|p{1.6cm}|p{1.5cm}|p{1.5cm}|p{1.6cm}|}
\hline
  Name & Pos. (m) & Type & Slits & Slit angle$^{\circ}$ & Max.Freq. (Hz) \\\hline \hline
PS   & 9.5   &  2 x CR  & 2      & 50 &  300 \\
 RRM                      & 19  & CR    & 2     & 4.88  & 300 \\
  BW        & 19.05    & SD             & 1     & 4.88   & 14 \\
   F0       &  27.5      & SD             & 1    & 110   & 250 \\
 F0     & 27.55          & SD             & 1    & 110   & 250\\
 M  & 28.5 & 2 x CR     & 1   & 30 &   400 \\
    \hline
\end{tabular}
\caption{Chopper parameters. CR = Counter rotating, SD = single disk}
\label{Chopper}
\end{minipage}
\end{table}

\section{Background considerations.}
Modern spallation neutron sources are driven by proton beams with GeV energies. Whereas low energy particle background shielding is well understood for reactors sources of neutrons (~20 MeV), for high energies (100s MeV to multiple GeV) there is 
potential to improve shielding solutions and reduce instrument backgrounds significantly.  This is particularly pertinent for the more recent spallation sources that have come online  with 0.91 GeV for the SNS, 3.0 GeV for J-Parc and 2.5 GeV for ESS. The principal source of background originates from the prompt pulse of protons on the target that translates into a long time dependent tail of fast neutron spectra at the sample. 
A prime candidate mechanism in the prompt pulse is the phenomenon of particle showers. These are the cascades of secondary particles produced by high energy particle interactions with dense matter. Chopper spectrometers are particularly sensitive to these time dependent particle showers with tails extending up to 3~ms. Chopper spectrometers minimise the problem through the use of curved guides and or T0 choppers that absorb the high energy background. Indeed, 4Season of  J-Parc, a thermal chopper spectrometer with a distance of 18~m between moderator to sample, recently implemented an optimised T0 chopper that was able to reduce the background by three orders of magnitude \cite{JParcT0}. The background on VOR will be limited by a combination of a curved guide and T0 choppers. The guide is curved to ensure that no direct line of sight is possible between the moderator and the sample thus limiting a substantial contribution to the background. The precise details of the T0 chopper design has not yet been finalised since this requires an understanding of the target and moderator concepts that have not yet been achieved. As such the details of a T0 chopper is not included in Figure \ref{ChopCascade}. 

\subsection{Flux at the sample}
A lack of flux is commonly cited as the limitation to accessing novel information in neutron scattering.  Indeed it can be expected that an incremental increase in scientific knowledge will be gained by a factor of ten improvement in neutron flux while entirely  new scientific domains will be accessed with two orders of magnitude.   \\
\begin{figure}
\centering
\includegraphics[width=0.45\textwidth]{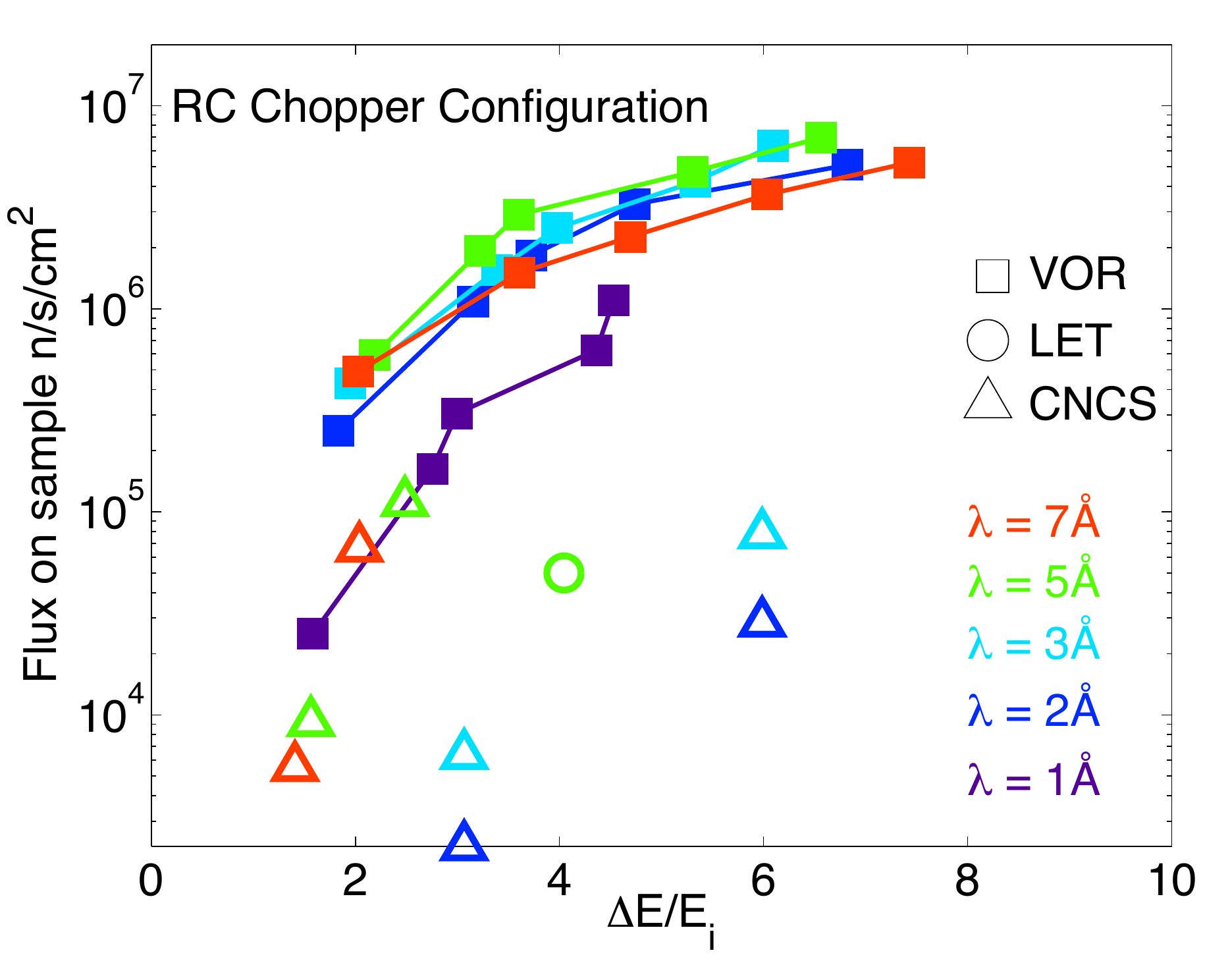}
\caption{Neutron flux on sample as a function of energy resolution for a range of incident wavelengths for the relative constant energy resolution chopper configuration.}
\label{Fluxes}
\end{figure}
The fluxes expected on VOR are given in Figure \ref{Fluxes}.  The results are compared to flux data provided by CNCS at the SNS \cite{CNCS} and LET \cite{LET} at ISIS. The VOR Chopper Spectrometer can offer a ten-fold increase in flux at equivalent resolution over the entire wavelength range for a single incident wavelength. However, with a marginal relaxation of resolution, $\Delta$ E/E = 2 $\rightarrow$ 5 \%, the VOR chopper spectrometer is able to offer an increase of two orders of magnitude in flux for each incident wavelength in comparison to existing instruments. It should be noted that Figure \ref{Fluxes} does not take into account multiple pulses provided by RRM that will increase the gains of VOR by a further order of magnitude for each ESS time pulse.

\subsection{Detectors}
The sample to detector distance on VOR is 3~m and  provides a $\Delta E/E$ of 1~\% for the highest wavelengths whilst maintaining a reasonable energy resolutions for the lowest wavelengths, considering a distance uncertainty of 2~cm that includes the sample and detector depths. Detector technology will be based on $^{10}$B thin film technology. Current state-of-the-art PSD arrays in neutron scattering employ $^{3}$He tubes as the detector technology of choice. However $^{3}$He is no longer available for large detector areas \cite{helium-3-crisis_a, helium-3-crisis_b, helium-3-crisis_c, Kouzes}. The detector array envisaged on VOR will employ the multi-grid detector design ~\cite{AndersenDet} \cite{multigrid} which incorporates thin conversion layers of 1~$\mu$m thick B$_{4}$C Carbide~\cite{B4C} coupled to segmented square tubes incorporating an anode wire and counting gas for signal detection.   This design foresees a 1~cm$^3$ voxel design; however it should be noted that the depth, hence time resolution of such a device can be made extremely good if required, due to the conversion in a very thin layer~\cite{res-depth}. The macrostructured $^{10}$B multi-wire proportional chamber comprises the backup detector option~\cite{b10-data_d}. Optimisation of the design parameters will be needed to achieve optimal performance, however these are well understood analytically~\cite{b10-analytical_a, b10-analytical_b} and have been recently verified~\cite{b10-verif_a,b10-verif_b}.The detector tank is an integral part of the structure for the multigrid design. Windows in the detector tank will be minimised with the guide exit incorporated into the detector tank. The main volume of the detector tank will be pumped to cryogenic vacuum, 10$^{-6}$ mbar.  The entire tank will be non-magnetic to enable polarisation analysis and high magnetic field measurements. The detector bank covers scattering angles from -40 to 140~$^{\circ}$ in the horizontal plane $\pm$27$^{\circ}$ in the vertical plane, a detector area of approximately 30~m$^{2}$. A long "get-lost" tube and an extended beamstop limits the background contributions at low scattering angles.\\

\section{Conclusion}
A design study of the versatile optimal resolution chopper spectrometer for the ESS, VOR, is presented. The instrument is optimised for the study of small sample volumes, weak scattering features and  transient phenomena that can be accessed via 
enormous flux gains on this wide bandwidth instrument. The flux gains, in conjunction with the novel RC chopper configuration, are expected to hugely enhance the entire field of neutron scattering by  accessing scientific domains that are currently out of reach.  
\section{Acknowledgements}
The support from and useful discussions with the Scientific and Technical Advisory Panel (STAP) for Direct Geometry Spectroscopy of the ESS is gratefully acknowledged.



%
%
\bibliography{Deen_P_qenswins2014.bib}


\end{document}